\documentclass[aps,prb,reprint,numerical]{revtex4-1}

\usepackage{hyperref}
\usepackage{amsmath}
\usepackage{amsfonts}
\usepackage{amssymb}
\usepackage{graphicx}

\begin{document}
\title{Low-power photothermal probing of single plasmonic nanostructures with nanomechanical string resonators}

\author{Silvan Schmid}
\email{sils@nanotech.dtu.dk}
\author{Kaiyu Wu}
\author{Peter Emil Larsen}
\author{Tomas Rindzevicius}
\author{Anja Boisen}
\affiliation{Department of Micro- and Nanotechnology, Technical University of Denmark, DTU Nanotech, DK-2800 Kgs. Lyngby, Denmark}




\begin{abstract}
We demonstrate the direct photothermal probing and mapping of single plasmonic nanostructures via the temperature induced detuning of nanomechanical string resonators. Single Au nanoslits are illuminated with a low-power polarized focused laser beam ($\lambda = 633$~nm). Polarization dependent heat generation in gold nanoslits is then imaged with high sensitivity. A sensitivity of -4.1~ppm/nW with respect to the illuminated light (beam diameter $5.0\pm0.8$~$\mu$m) is determined for a single nanoslit (1~$\mu$m long and 53~nm wide), which equals to a total light absorption of 16\%. This results in a heating of 0.5~K for an illuminance of 8~nW/$\mu$m$^2$. Our results show that nanomechanical resonators are a unique and robust analysis tool for the low-power investigation of thermoplasmonic effects in plasmonic hot spots.
\end{abstract}

\maketitle
Sub-wavelength noble metal structures (nanoparticles or nanovoids) support localized surface plasmon (LSP) resonances that usually occur in the visible and near-infrared spectral region. The incident light can couple to LSP modes producing extremely large field enhancements, so-called hotspots. These strong field confinements are prominently utilized e.g. in surface enhanced Raman scattering (SERS) spectroscopy, \cite{Nie1997,Kneipp1997,Haynes2005,Moskovits2005} in plasmonic solar cells, \cite{Atwater2010,Pillai2007,Catchpole2008,Ferry2010,Derkacs2006} or as nano heat sources with a wide range of potential applications. \cite{Baffou2013,Baffou2010a,Schuller2010}
When a plasmonic nanostructure is illuminated with an incident light, part of it is scattered into the surrounding medium, while the other part is absorbed and dissipated as heat. Interestingly, optical and thermal hot spots are generally mismatched. G. Baffou et al. have shown experimentally that heat is concentrated in areas where charges can freely flow, while optical hot spots usually appear at the metal interface with the greatest charge accumulation (tip effect).\cite{Baffou2013} So far there have been a few attempts to investigate the photothermal heating of metal nanostructures using thermal optical techniques \cite{Baffou2010} and an AFM tip. \cite{Lahiri2013} However, the desirable direct investigation of the heating mechanisms in plasmonic structures is still a challenge due to the lack of robust experimental tools. \cite{Baffou2013}
Herein we propose a new approach to probe and image plasmonic structures with nanoscale resolution by measuring the photothermally induced frequency detuning of highly temperature sensitive nanomechanical resonators. It has been shown that plasmons can be coupled to the vibration of a nanomechanical resonator. \cite{Thijssen2013} We employ the high temperature sensitivity of a nanomechanical string resonator \cite{Larsen2013,Larsen2011} to directly probe the heating pattern produced by single gold nanoslits illuminated by a polarized scanning laser beam. The experimental approach allows a sensitive heat mapping of nobel metal nanostructures to study heat generation and thermal diffusion in e.g. plasmonic hot spots.

\begin{figure}
  \includegraphics[width=8.46cm]{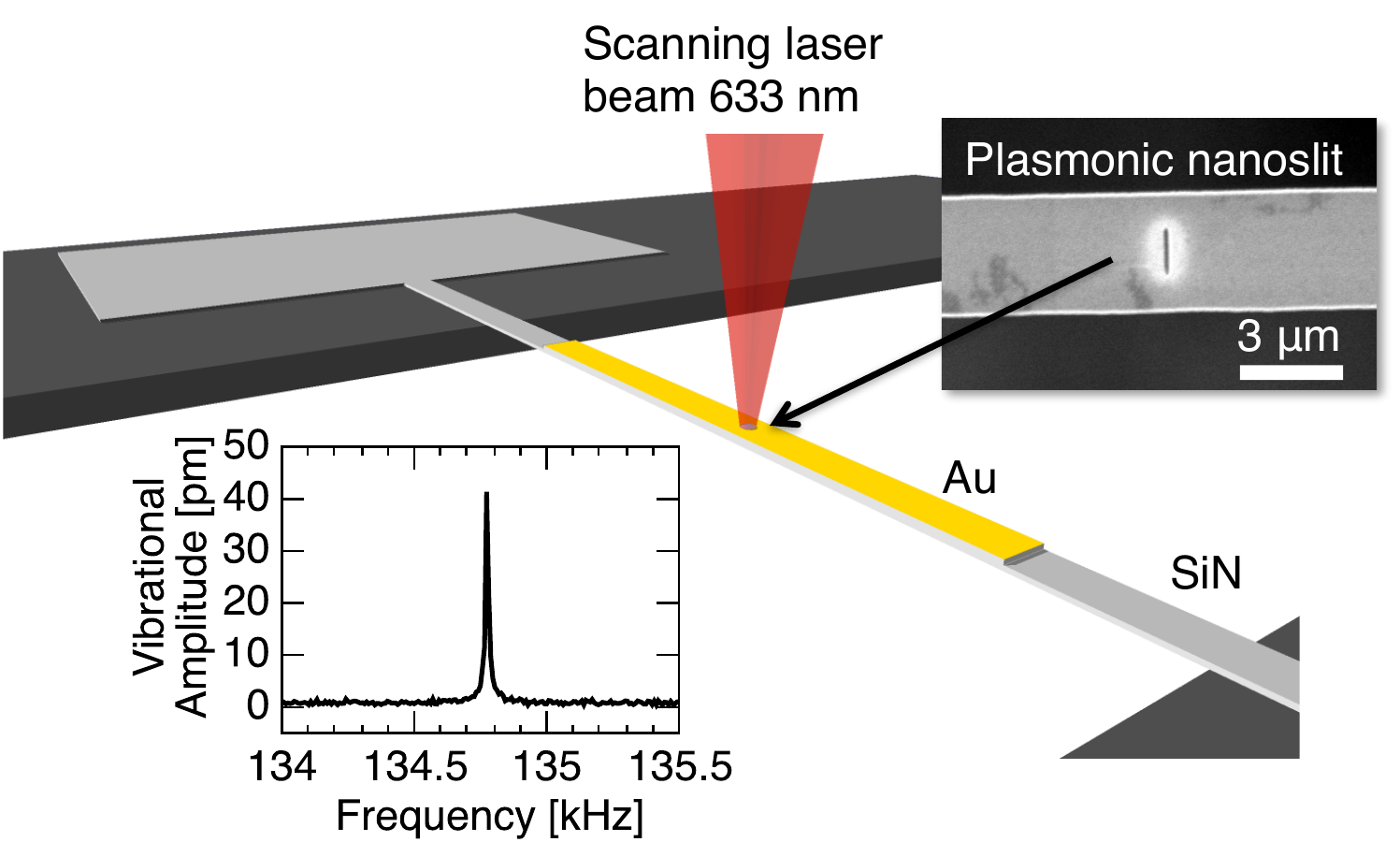}
  \caption{Schematic depiction of the experimental setup. A nanomechanical SiN string resonator is partially coated with a Au layer. A plasmonic nanoslit (etched with a focused ion beam) is located in the string center and probed with a low-power focused laser beam from a laser-Doppler vibrometer. The inset shows the thermal fluctuation spectrum of a 900~$\mu$m long micromechanical SiN string resonator.}
  \label{fig:1}
\end{figure}

The proposed method to photothermally probe plasmonic nanostructures is schematically illustrated in \autoref{fig:1}. Single Au nanoslits were chosen as a nano heat source. The nanoslits were milled using a focused ion beam (FIB) (using a 30~keV Ga+ beam with current of 9~pA) into the Au metal coating of a nanomechanical silicon nitride (SiN) string resonator, and the separation between individual nanoslits is $\sim 10$~$\mu$m. The fabrication of the SiN string resonators is described elsewhere \cite{Schmid2011}. The string resonance frequency corresponds to the thermal vibration of the string (see inset of \autoref{fig:1}), measured in high vacuum with a laser-Doppler vibrometer (MSA-5 from Polytec). The s or p-polarized vibrometer's laser beam is utilized to simultaneously record the string resonance frequency and excite localized and surface plasmons in individual Au nanoslits. Part of the incident light is absorbed by a nanostructure and dissipated into heat. The string resonator is highly temperature sensitive and can be used to monitor the photothermal heating produced by the nano heat source on the string via the resonance frequency detuning.

The relative resonance frequency change of a string resonator as a function of absorbed power $P$ in the center can be calculated from\cite{Yamada2013}
\begin{equation}\label{eq:dfvsP}
\delta f = \frac{\Delta f}{f} = - \frac{L}{16 w h}\frac{\alpha E}{\sigma \kappa}P,
\end{equation}
with the string length $L$, thickness $h$, width $w$, thermal expansion coefficient $\alpha$, Young's modulus $E$, tensile pre-stress $\sigma$, and thermal conductivity $\kappa$. In the case of a bilayer string, the material parameters can be replaced by the effective parameters based on the arithmetic average with respect to the layer thicknesses.

\begin{figure}
  \includegraphics[width=8.46cm]{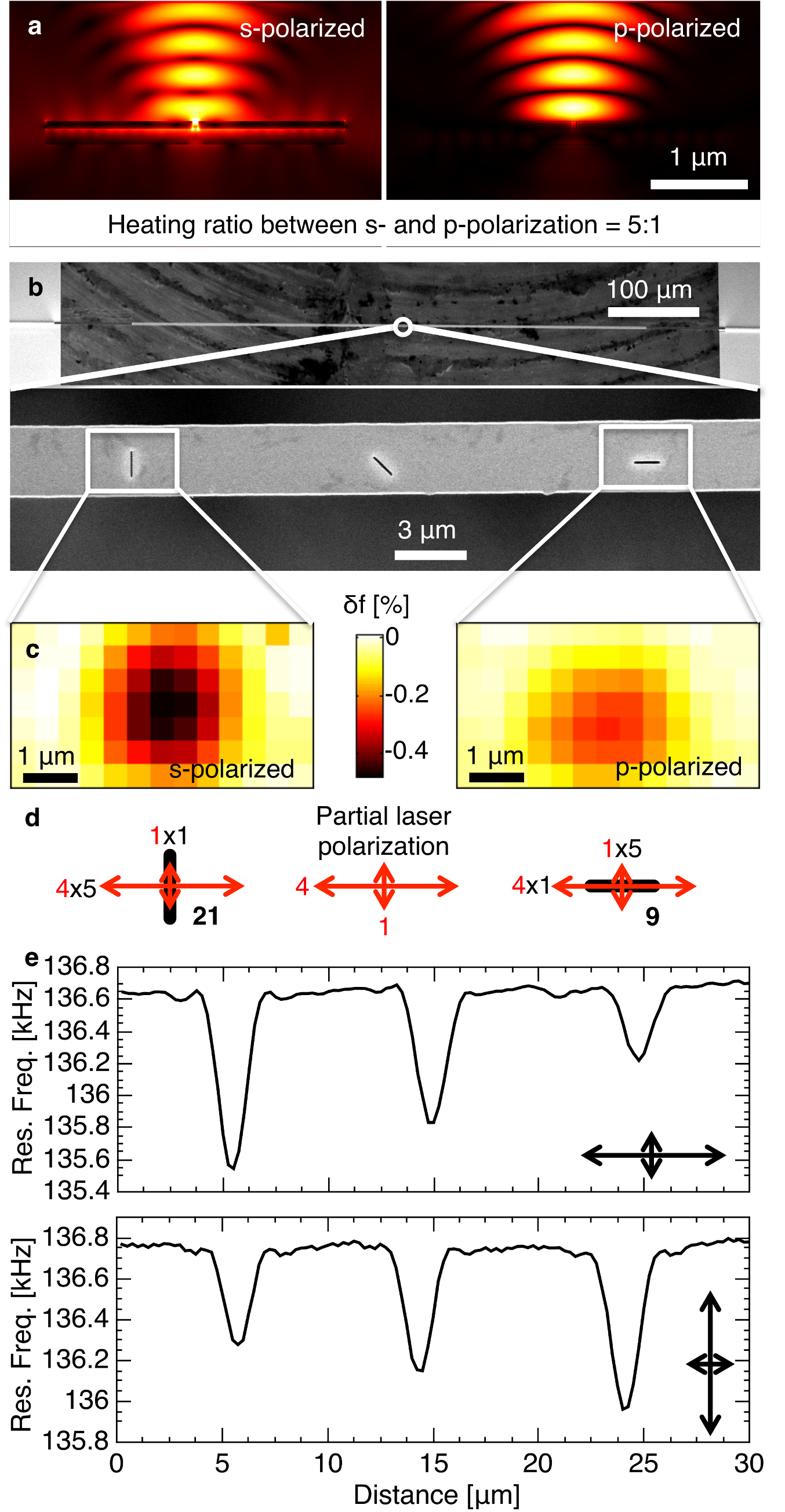}
  \caption{(a) FEM simulation of the modulus amplitude of the electric field (ranged from 0~V/m - 2~V/m) for a SiN string (157~nm thick and 3~$\mu$m wide) covered with 80~nm Au, illuminated with a $\lambda = 633$~nm laser. (b) SEM image of a Au coated stoichiometric SiN string ($\sigma=900$~MPa, $L=900$~$\mu$m, $w=3$~$\mu$m, $h=157$~nm, covered with 80~nm Au), featuring plasmonic nanoslits in the center (90~nm wide and 1~$\mu$m long). (c) Photothermal maps of a vertical and horizontal plasmonic slit with a partially horizontally polarized laser beam with a diameter of $\sim 1-2$~$\mu$m and a power of $P=600$~nW. (d) Schematic calculation of the heating ratio between a vertical and horizontal nanoslit. (e) Resonance frequency of the string for a scan over all three plasmonic nanoslits for two perpendicular laser orientations with a power of $P=1350$~nW.}
  \label{fig:2}
\end{figure}


The absolute value of the electric field around a 2-dimensional (infinitely long) Au nanoslit was simulated using the finite element method (FEM). \autoref{fig:2}a shows the $|E|$ map in a string cross-section for s-polarized (perpendicular) and p-polarized (parallel) incident light with respect to the nanoslit orientation. In the s-polarized case the incoming field is enhanced around the slit due to the localized nature of the plasmonic excitation. Additionally, a surface plasmon polariton (SPP) is launched at the metal-insulator interface. In the p-polarized mode, the LSP generation is minimized. The resistive loss and thus the heating in the metal of a string is roughly 5 times larger for s-polarization compared to p-polarization. The polarization-dependent photoinduced heating of the string resonator is tested using three identical nanoslits with different orientations as shown in \autoref{fig:2}b. Photothermal maps around vertically and horizontally aligned single nanoslits are shown in \autoref{fig:2}c, and the spatial resolution is $\sim 375$~nm. According to \autoref{eq:dfvsP}, the thermal heating is directly proportional to the measured relative frequency shift ($\Delta f/f$) of the string resonator. In the s-polarization case, the produced frequency shift, and thus the heat, is roughly double in magnitude compared to the p-polarization case. The laser beam is partially polarized with a power ratio of 4:1 in the two directions of polarization. With the  heating ratio 5:1 of s- vs. p-polarization, all heating contributions from both polarization directions add up to a total heating ratio of 21:9 (see \autoref{fig:2}d). This is well in agreement with the measured double frequency detuning of s-polarization compared to p-polarization. A line scan over all three slits is depicted in \autoref{fig:2}e, showing the continuous polarization-dependent heating for three different nanoslit orientations. The observed polarization dependence is clear evidence of the plasmonic nature of the absorption enhancement of the Au nanoslits.

\begin{figure}
  \includegraphics[width=7cm]{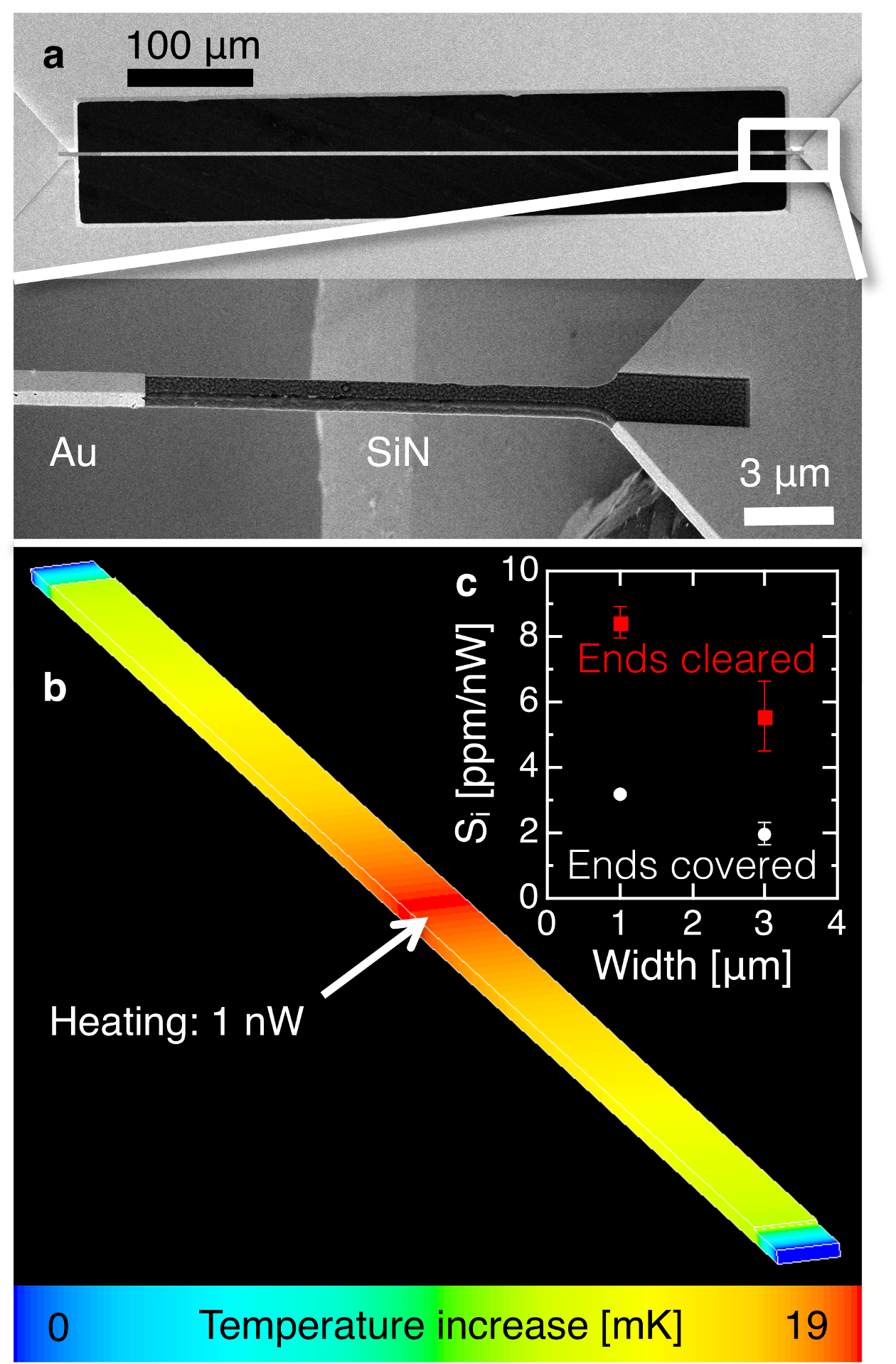}
  \caption{(a) SEM image of a silicon-rich SiN nanostring ($\sigma=200$~MPa, $L=550$~$\mu$m, $h=220$~nm, covered with 50~nm Au) and a zoom in to the anchor area showing the by FIB removed Au layer ($2\times 2.5$\% of string length). (b) FEM simulation of the temperature field in the nanostring due to a heating power of 1~nW in the string center. (c) Sensitivity with respect to the illuminated laser power measured with the silicon-rich SiN nanostrings with and without Au covered ends. The error bars represent the standard deviation of 5 measurements.}
  \label{fig:3}
\end{figure}

In order to use a nanomechanical string resonator as a quantitative probe, its sensitivity has to be evaluated. Based on \autoref{eq:dfvsP}, the sensitivity with respect to absorbed power $S_a = \delta f/P$ is defined. In the measurements, only the absolute illuminated laser power is known. We therefore further define the sensitivity with respect to the illuminated power $S_i = \alpha S_a$, with the factor $\alpha$ being the ratio between absorbed to illuminated power.

Based on \autoref{eq:dfvsP}, respective absorption sensitivities $S_a=-9.6\pm2.1$~ppm/nW and $S_a=-3.2\pm0.6$~ppm/nW can be calculated for a 1~$\mu$m and 3~$\mu$m wide, fully covered SiN/Au bilayer string with a length of 550~$\mu$m, as shown in \autoref{fig:3}a, with $E_{SiN}=240$~GPa \cite{Schmid2011},$E_{Au}=78$~GPa, $\alpha_{SiN}=1.23$~ppm/K \cite{Larsen2011},$\alpha_{Au}=14$~ppm/K, $\kappa_{SiN}=2.5$~W/(m~K), and $\kappa_{Au}=320$~W/(m~K). In order to improve the sensitivity, the Au film at the anchors of the strings was removed by FIB milling. A close up of an anchor area is shown in \autoref{fig:3}a. The isolating effect of the gap in the Au layer is clearly visible in the simulated temperature field shown in \autoref{fig:3}b. By removing roughly 5\% of the Au layer, the illumination sensitivity $S_i$ of a 1~$\mu$m and 3~$\mu$m wide string could be improved by a factor of 2.6 and 2.8, respectively, as can be seen from the measurements presented in \autoref{fig:3}c. The gained enhancement of $S_i$ translates directly to an enhancement of the absorption sensitivity. Values of $S_a = -25.0\pm5.5$~ppm/nW and $S_a = -9.0\pm1.7$~ppm/nW can thus be estimated for 1~$\mu$m and 3~$\mu$m wide strings, respectively. These values agree well with respective sensitivities $S_a = -25.1\pm1.3$~ppm/nW and $S_a = -8.3\pm0.5$~ppm/nW calculated with FEM.


\begin{figure}
  \includegraphics[width=8.46cm]{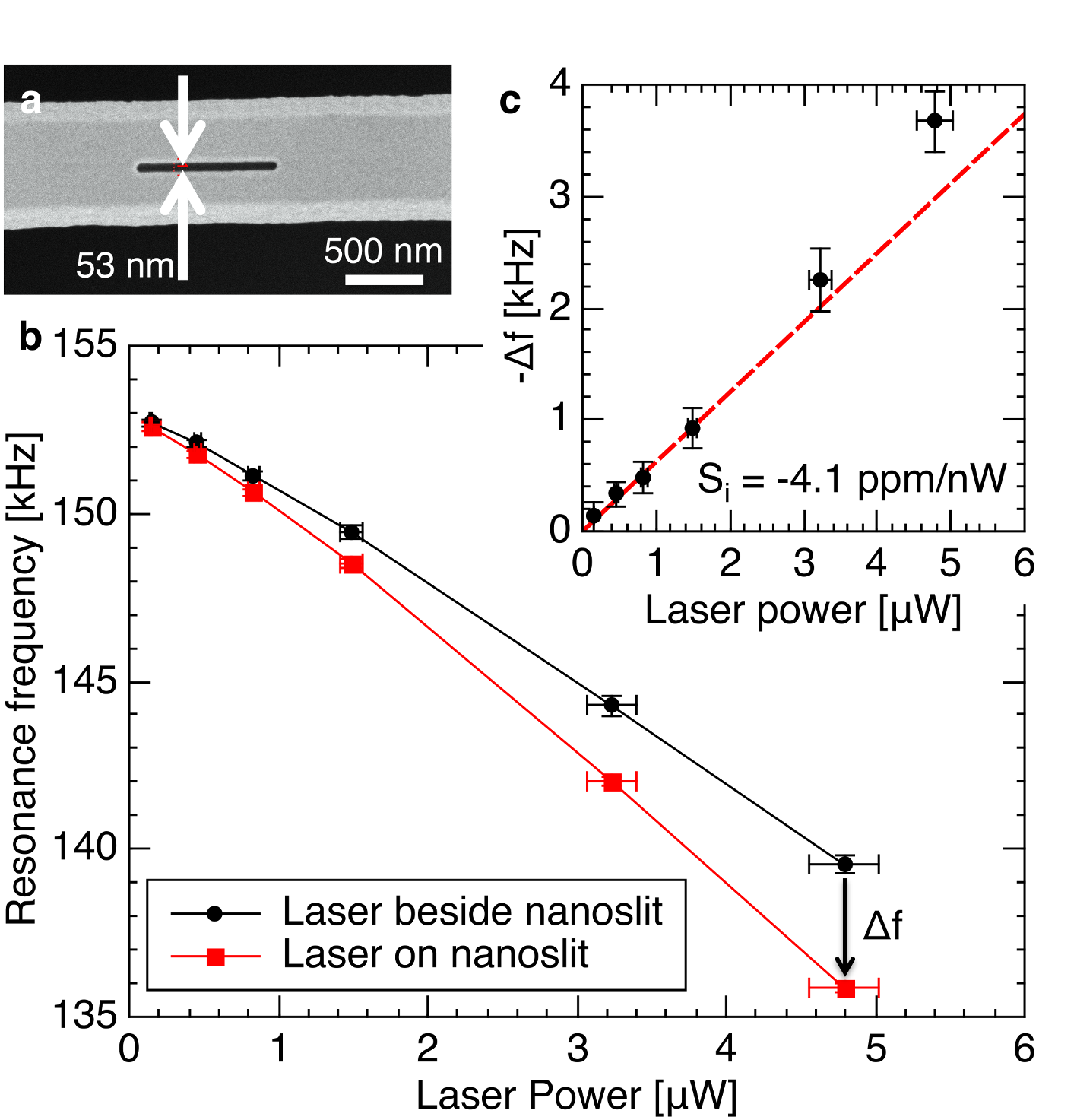}
  \caption{(a) SEM image of nanoslit in the center of a silicon-rich SiN nanostring ($\sigma=200$~MPa, $L=550$~$\mu$m, $w=1$~$\mu$m, $h=220$~nm, covered with 50~nm Au). (b) Measured resonance frequency of this nanostring as a function of illuminated laser power when focused on the plasmonic nanoslit and when focused beside the nanoslit. The laser beam diameter was approximately 5.0$\pm$0.8~$\mu$m. (c) Frequency difference caused by the photothermal heating of the plasmonic nanoslit. The error bars represent the standard deviation of 3 measurements.}
  \label{fig:4}
\end{figure}

\autoref{fig:4} shows the photothermal heating as a function of laser power of a single plasmonic nanoslit (see \autoref{fig:4}a) on a 1~$\mu$m wide nanomechanical string (equal to the string depicted in \autoref{fig:3}a and discussed in the previous paragraph). In this case the laser beam diameter is set to 5.0$\pm$0.8~$\mu$m in order to achieve more uniform illumination power across the string. \autoref{fig:4}b shows the resonance frequency of the string when focusing the laser beam either directly on the Au nanoslit or planar gold film. The frequency difference, as shown in \autoref{fig:4}c, is directly caused by the additional photothermal heating of the plasmonic nanoslit. From the slope at low laser powers, an illumination sensitivity of $S_i=-4.1$~ppm/nW can be extracted. At higher powers, the frequency detuning becomes slightly nonlinear. Considering the estimated absorption sensitivity of $S_a=-25.0$~ppm/nW, this corresponds to a ratio of absorbed to illuminated laser light of $\alpha = 16$~\%. That is, the 90~nm wide and 1~$\mu$m long plasmonic nanoslit absorbs 16~\% of the illuminated laser light with a beam diameter of 5.0$\pm$0.8~$\mu$m and a wavelength of $\lambda=633$~nm. With a simulated temperature sensitivity of 19~mK/nW (see \autoref{fig:3}b), this results in a nanoslit induced heating of 0.5~K and 14.6~K for the lowest (147~nW) and highest (4.8~$\mu$W) laser power, respectively. The lowest illuminance used in this experiment of 8~nW/$\mu$m$^2$ is more than 6 orders of magnitude lower than the illuminance of 40~mW/$\mu$m$^2$ typically used in state-of-the-art probing of plasmonic nanostructures. \cite{Baffou2010} Low illuminances are critical for studying thermal hot spots with (bio)-molecules adsorbed on the metal surface to avoid photochemistry related effects or thermal decomposition of adsorbates.

In conclusion, we demonstrate the low-power photothermal probing and mapping of single plasmonic Au nanoslits via the temperature-induced detuning of the resonance frequency of nanomechanical SiN string resonators. The polarization dependent heating patterns are mainly observed in the vicinity of the nanoslits that can be attributed to excited LSPs and the SPPs at the Au/SiN interface. We produced a photothermal map with a resolution of $\sim 375$~nm by scanning the focused probing laser over the nanoslit area. The photothermal sensitivity of the string resonators was improved by a factor of $\sim 3$ by removing 5\% of the thermally highly conductive Au layer at the string ends. Thereby, an absorption sensitivity for a nanomechanical silicon-rich SiN string ($L=550$~$\mu$m, $w=1$~$\mu$m, $h=220$~nm, covered with 50~nm Au) of $S_a=-25.0\pm5.5$~ppm/nW was estimated based on an analytical model and confirmed using FEM. The measured illumination sensitivity of $S_a=-4.1$~ppm/nW of a single nanoslit (1~$\mu$m long and 53~nm wide) on such a string leads to an absorption ratio of $\alpha=16$\% of a laser beam with a diameter of 5.0$\pm$0.8~$\mu$m. This corresponds to a temperature increase of 0.5~K for a laser power of 147~nW. Our results show that nanomechanical resonators are a unique and robust tool for probing thermal effects in plasmonic nanostructures.

\section*{Acknowledgment}
This research is supported by the Villum Foundation's Young Investigator Program (Project No. VKR023125). The authors thank Zolt\'an Imre Balogh and Adam Fuller for the support with the focused ion beam milling, and Jens Q. Adolphsen for the cleanroom support.


%

\end{document}